\RequirePackage{lineno}
\documentclass[twocolumn,aps,prb,superscriptaddress,showpacs]{revtex4}
\usepackage{graphicx}
\usepackage{amsmath}
\usepackage{amsfonts}
\usepackage{color}
\usepackage{ulem}
\usepackage{epstopdf}
\usepackage{sidecap}
\usepackage{dsfont}

\sidecaptionvpos{figure}{c}

\begin{document}


\title{Disorder effects in multiorbital $s_{\pm}$-wave superconductors: Implications for Zn-doped BaFe$_2$As$_2$ compounds}

\author{Hua Chen}
\affiliation{Zhejiang Institute of Modern Physics and Department of Physics, Zhejiang University, Hangzhou 310027, China}

\author{Yuan-Yen Tai}
\affiliation{Department of Physics and Texas Center for Superconductivity, University of Houston, Houston, Texas 77204, USA}
\affiliation{Theoretical Division and Center for Nonlinear Studies, Los Alamos National Laboratory, Los Alamos, New Mexico 87545, USA}

\author{C. S. Ting}
\affiliation{Department of Physics and Texas Center for Superconductivity, University of Houston, Houston, Texas 77204, USA}

\author{Matthias J. Graf} 
\affiliation{Theoretical Division and Center for Nonlinear Studies, Los Alamos National Laboratory, Los Alamos, New Mexico 87545, USA}

\author{Jianhui Dai}
\affiliation{Condensed Matter Group, Department of Physics, Hangzhou Normal University, Hangzhou 310036, China}
\affiliation{Zhejiang Institute of Modern Physics and Department of Physics, Zhejiang University, Hangzhou 310027, China}

\author{Jian-Xin Zhu}
\email[To whom correspondence should be addressed.\\]{jxzhu@lanl.gov}
\homepage{http://theory.lanl.gov}
\affiliation{Theoretical Division and Center for Nonlinear Studies, Los Alamos National Laboratory, Los Alamos, New Mexico 87545, USA}
\affiliation{Center for Integrated Nanotechnologies, Los Alamos National Laboratory, Los Alamos, New Mexico 87545, USA}

\date{\today}

\begin{abstract}
Recent experiments on Zn-doped 122-type iron pnictides, Ba(Fe$_{1-x-y}$Co$_y$Zn$_x$)$_2$As$_2$, are challenging our understanding of electron doping the 122s and the interplay between doping and impurity scattering.
To resolve this enigma, we investigate the disorder effects of nonmagnetic Zn impurities
in the strong (unitary) scattering limit on various properties of the system in the $s_{\pm}$-wave superconducting pairing state. 
The lattice Bogoliubov-de Gennes equation (BdG) is solved self-consistently based
on a minimal two-orbital model with an extended range of  impurity concentrations.
We find that Zn impurity is best modeled as a defect, where charge is mainly localized, but scattering is extended over a few lattice sites.
With increasing Zn concentration the density of states shows a gradual filling of the gap, revealing the impurity-induced pair
breaking effect. Moreover, both the disorder configuration-averaged superconducting order parameter and the superfluid
density are dramatically suppressed towards the dirty limit,
indicating the violation of the Anderson theorem for conventional $s$-wave superconductors and the breakdown of the Abrikosov-Gorkov theory for impurity-averaged Green's functions. Furthermore, we find that the superconducting phase is fully suppressed 
close to the critical impurity concentration of roughly $n_\text{imp}\approx 10\%$, in agreement with recent experiments.
\end{abstract}

\maketitle

\section{Introduction}
The superconductivity in iron pnictides has received tremendous
interest since its discovery.~\cite{kamihara08,chenxh08,chengf08,ren08,wang08} 
The generic phase diagram of the iron-based superconductors (Fe-SCs)
suggests the close proximity of the superconductivity (SC) to the spin-density-wave (SDW) 
antiferromagnetism (AFM).~\cite{pcdai08} 
This should be contrasted with
high-temperature cuprates, of which the SC originates from an AFM Mott insulator phase and 
can be quickly suppressed by substitution with other
$3d$-transition-metal atoms into the CuO$_2$ plane.
Instead, the SC in the Fe-SCs  can be
induced when Fe atoms are partially replaced by $3d$-transition-metal atoms like
Ni and Co.~\cite{rotter08,ALeithe-Jasper:2008,NNi:2008,HQLuo:2008,GFChen:2008,ASSefat:2008,LJLi:2009,SRSaha:2009}
Moreover, in the case of the electron- and hole-doped 122 family
like Ba(Fe$_{1-x}$Co$_x$)$_2$As$_2$ or Ba$_{1-x}$K$_x$Fe$_2$As$_2$,
the coexistence of the SDW and SC in a narrow doping region is
reported by both  experiments~\cite{gordon10,rotter08} and theoretical calculations.~\cite{MGVavilov:2009,TZhou:2011} 
It has been argued that owing to their
multiorbital nature and variable correlation effects, the superconducting  pairing
symmetry may not be universal in the Fe-SCs.~\cite{si09,hirschfeld11,hu12} 
This certainly poses a great challenge to relating the symmetry of the order parameter and the canonical doping phase diagram across different crystallographic iron-pnictide families.
Therefore, detailed measurements of the bulk transport and superconducting properties will remain useful for determining the superconducting pairing symmetry in the Fe-SCs.

While superconductivity can be induced in the 122 family by substitution of Fe with Co or Ni (and many other $3d$ transition metals~\cite{SBZhang:2011,YQi:2011,YNishikubo:2010,FHan:2009}) in the
FeAs layer, the role of these electron dopants is still
controversial.~\cite{onari09,boyd10,efremov11,fernandes12,wang13}
It is hotly debated whether such substitution is dominated by the doping effect of extra
charge carriers or impurity scattering or a combination thereof.~\cite{berlijn12} 
In general, the study of disorder effects in superconductors is promising in addressing the pairing symmetry, as it has been applied successfully to the understanding of high-$T_c$ cuprate superconductivity.~\cite{AVBalatsky:2006}  Therefore, it is natural that the study of impurity scattering effects in Fe-SCs attracted much attention immediately after their discovery.   Of particular interest is the case of Zn doping,~\cite{li09,li10,yao12}  because it does not induce SC, while it is expected to
give rise to nonmagnetic, strong potential scattering in the unitary limit. For the Zn-doped 122-type iron pnictides, the
early results showed that  Zn impurities hardly affect the
superconductivity of hole-doped
Ba$_{0.5}$K$_{0.5}$Fe$_2$As$_2$.~\cite{cheng10} 
However, more recent measurements of the
magnetic susceptibility and resistivity~\cite{li11,li12a} on high-quality single-crystalline
Ba(Fe$_{1-x-y}$Zn$_x$Co$_y$)$_2$As$_2$ compounds, suggested that the electron doped
superconductivity is almost fully suppressed above a concentration of roughly $8\%$ Zn, regardless of
whether the sample is under, optimally or over doped.  This
discrepancy with earlier experiments is possibly due to the technical difficulty in substituting
Zn for Fe atoms. Further measurements~\cite{li12b} on the hole doped
Ba$_{0.5}$K$_{0.5}$(Fe$_{1-x}$Zn$_x$)$_2$As$_2$ compound also showed that the
superconductivity is suppressed by Zn impurities.  These interesting results have presented a challenge to theoretically identify the pairing symmetry in Fe-SCs. 
So far, the
sign-reversal $s_{\pm}$-wave pairing symmetry has been supported by many experiments including neutron scattering,~\cite{christianson08} angle-resolved photoemission spectroscopy,~\cite{HDing:2008}  
 and scanning tunneling spectroscopy,~\cite{hanaguri10}  and is also consistent with the competition picture between magnetism and superconductivity.~\cite{fernandes10} 
However, it has also been shown earlier~\cite{onari09} that because of the sign reversal of superconducting gap function  across electron and hole bands, the $s_\pm$-wave pairing state is very fragile against impurities while the non-sign-reversal $s_{++}$-wave pairing symmetry should be a competitive candidate for Fe-SCs. The recent experiment~\cite{li12a} showed  that the suppression in the superconducting transition temperature is much slower than that predicated by the theory for the $s_{\pm}$-wave pairing state.~\cite{onari09} 
More recently, the effect of
Zn-doping induced disorder in Fe-SCs with
both $s_{\pm}$- and $s_{++}$-wave pairing symmetries~\cite{yao12} has been
investigated by solving the BdG equation for a
two-orbital model~\cite{SRaghu:2008}  including both on-site (favoring $s_{++}$-wave pairing symmetry)
and next nearest neighbor (NNN) inter-site (favoring $s_{\pm}$-wave pairing symmetry) pairing
interaction.  The zero-temperature real-space BdG calculations~\cite{yao12} indicated that the disorder could suppress the NNN pairing order parameter with negligible effect on the on-site pairing order parameter, suggesting a possibility of disorder induced pairing symmetry change from $s_{\pm}$- to $s_{++}$-wave.  As such, depending on the strength of the on-site pairing interaction, this interesting proposal may provide a flexibility to explain various experimental data.~\cite{li09,li10,yao12} 

We note that, in Ref.~\onlinecite{yao12}, because the tuning of  impurity concentration in the truly disordered system was actually mimicked by a tuning of the NNN pairing interaction in an impurity-free system, a direct comparison of superconducting transition temperature change with impurity concentration between theory and experiment is impossible. Due to this interpretation gap, there are still several open questions.
In this paper, we will study the disorder effects of the Zn impurity on the superconducting properties of 122-type  iron-based superconductors. In particular, we aim to address the question of how the superconducting transition temperature is completely suppressed at 8\% of Zn doping in 122-type compounds.~\cite{li11,li12a,li12b}  To fulfill this goal, we start with an improved minimal two-orbital model for Fe-SCs.~\cite{YYTai:2013} As in Ref.~\onlinecite{yao12}, we solve the BdG equations self-consistently in real space to study the impurity-induced disorder effect,  from which the superconducting order parameter, superconducting transition temperature, superfluid stiffness are calculated. We point out that with the sole $s_{\pm}$-wave pairing symmetry, the superconducting transition temperature can be suppressed at an impurity concentration as high as about 10\%, which agrees well with the experiments on 
the  Ba(Fe$_{1-x-y}$Zn$_x$Co$_y$)$_2$As$_2$ compounds.~\cite{li12a} 
This result is in striking contrast with an earlier prediction that the superconductivity is suppressed already at only 1\% of impurity concentration.~\cite{onari09} The root cause for this difference is given as follows: 
Firstly, first-principles electronic structure calculations suggest that substitution of the
nonmagnetic Zn atom in the iron-based 122 superconductors, pushes the
Zn-$3d$ impurity level considerably far below the Fe-$3d$ level, namely by about $\sim 8-10$ eV.\cite{wadati10,Nakamura11, berlijn12}  
Hence Zn substitution should be regarded as a strongly localized defect in the strong scattering (unitary) limit. Such strong potential scattering is
supported by more recent angle-resolved photoemission spectroscopy
measurements on Ba(Fe$_{1-x}$Zn$_x$)$_2$As$_2$.~\cite{ideta13a, ideta13b} Secondly, as shown later by our calculations, the superconducting coherence length can be very short, which is consistent with the experimental observation that Fe-SCs are extremely type-II superconductors with Ginzburg-Landau parameter as large as 250.~\cite{ASSefat:2008,MKano:2009}
In such a case, the applicability of the conventional approach based on the Abrikosov-Gorkov (AG)
pair-breaking theory in dilute alloys,~\cite{abrikosov61} which assumes a spatially
uniform suppression of the impurity-averaged order parameter and Green's functions, is in question. The failure of the AG theory to address consistently the superconducting and transport properties in high-temperature cuprate and some heavy-fermion superconductors with short coherence length is well documented.~\cite{MFranz:1997,TDas:2011}
Therefore, in order to go beyond the applicability of the early theoretical studies and to
reveal the interesting physics of highly disordered or {\it dirty} high-temperature iron-based superconductors, we
shall study the nonmagnetic impurity-induced disorder effects in the unitary limit of multiorbital superconductors by solving the lattice BdG
equation. This approach has proven to be quite successful in providing a consistent picture for the suppression of superconducting transition temperature and superfluid density in the inhomogeneous high-temperature cuprate and plutonium-based heavy-fermion superconductors.\cite{MFranz:1997,TDas:2011,KOhishi:2007}
In this paper, we emphasize the key role of strong electronic inhomogeneity induced by Zn substitution and how it could be probed in
the 122 iron pnictides.

The remainder of this paper is organized as follows: In
Sec.~\ref{sec:model} we introduce the model Hamiltonian and the
formalism. To set the stage for the highly disordered materials, the single impurity problem is briefly revisited in
Sec.~\ref{sec:single}. The disorder effects of the strong scattering
limit on the superconducting order parameter are discussed in
Sec.~\ref{sec:superconductivity}. In Sec.~\ref{sec:dos}, disorder effects on the local
density of states and the superfluid density or magnetic
penetration depth are discussed. Finally, a brief summary is given in
Sec.~\ref{sec:summary}.

\section{Model and formalism}
\label{sec:model} 

The multiorbital nature of iron-based superconductivity requires the construction of physically reliable and computationally efficient effective low-energy multiorbital models. In particular, a simple two-orbital model was first constructed by Raghu and co-workers.~\cite{SRaghu:2008} The Fermi surface topology resulting from this model captured well the shape reported by angle-resolved photoemission spectroscopy.~\cite{HDing:2008} However, it has some weaknesses in other aspects of the electronic band dispersion. For example, too much imbalance of Fermi velocities on the electron and hole bands  has been revealed in the study of the local electronic structure around a single impurity of an $s_\pm$-wave superconductor.~\cite{RBeaird:2012} 
Several groups~\cite{PALee:2008,CCao:2008,KKuroki:2008} have pointed
out that one needs at least three orbitals to accurately reproduce the electronic band structure calculated in the density functional theory within the local density approximation (LDA).
However, it has also been shown~\cite{SRaghu:2008,YRan:2009} that the other
Fe-3$d_{xz}$ and Fe-3$d_{yz}$ orbitals play an important role in the low-energy
physics of these materials. 
On the other hand, it has been argued that the canonical minimal model of the 122-type iron pnictides requires only two irons (2-Fe) with two orbitals, $d_{xz}$ and $d_{yz}$, per unit cell to account for the effects of the upper and lower As atoms with respect to the two-dimensional plane of the Fe square lattice.\cite{DZhang:2009,JHu:2012} It is worthy to mention that these 2-by-2-orbital models have successfully described the behavior of the collinear AFM and its competition with the superconducting order in the electron-doped part of the phase diagram. In very recent work, several of the present authors have improved the model original proposed in Ref.~\onlinecite{DZhang:2009} to give a unified description of the entire phase diagram covering both the electron- and hole-doped regimes.\cite{YYTai:2013} To our knowledge, this is the only 2-by-2-orbital model so far, in which the resultant low-energy electronic energy dispersion agrees well with LDA electronic structure calculations in the entire Brillouin zone of 122-type iron compounds.

Here we start with the improved 2-by-2-orbital model of Ref.~\onlinecite{YYTai:2013}. Interestingly, we wish to point out that this model of 2-by-2 orbitals per unit cell can be mapped exactly onto two decoupled one-site two-orbital models by recognizing a unitary rotation of orbitals between both Fe sublattices. The technical details of this mapping are given in the Appendix \ref{Appendix}.   We write the complete Hamiltonian for the two-dimensional Fe-square lattice as
\begin{equation}
\mathcal{H}=\mathcal{H}_0+\mathcal{H}_\text{I}+\mathcal{H}_\text{pair}+\mathcal{H}_\text{imp}\;.
\label{EQ:hamil_tot}
\end{equation}
Here $\mathcal{H}_0$ is the tight-binding Hamiltonian of the normal-state 
band structure describing hopping between Fe-$3d_{xz}$ and Fe-$3d_{yz}$ orbitals. The  lattice  Hamiltonian in the real space (see also Appendix~\ref{Appendix}) is given by 
\begin{equation}
\mathcal{H}_0=\sum_{\mathbf{i}\mathbf{j}\alpha\beta\sigma}t^{\alpha\beta}_{\mathbf{i}\mathbf{j}}d^{\dagger}_{\mathbf{i}\alpha\sigma}d_{\mathbf{j}\beta\sigma}-\sum_{\mathbf{i}\alpha\sigma}\mu
d^\dagger_{\mathbf{i}\alpha\sigma}d_{\mathbf{i}\alpha\sigma}\;,
\end{equation}
where
$d^\dagger_{\mathbf{i}\alpha\sigma}$ creates an electron with spin $\sigma$
in the effective orbitals $\alpha=1$ and $2$ on the
$\mathbf{i}$-th lattice site. We choose the nonvanishing hopping matrix
elements as $t^{\alpha\alpha}_{\pm \hat{x}}=t^{\alpha\alpha}_{\pm \hat{y}}=0.09$,
$t^{\alpha\bar{\alpha}}_{\pm \hat{x}}=t^{\alpha\bar{\alpha}}_{\pm \hat{y}}=-1$,
$t^{11}_{\pm (\hat{x} + \hat{y})}=t^{22}_{\pm (\hat{x}-\hat{y})}=1.35$,
$t^{11}_{\pm (\hat{x} - \hat{y})}=t^{22}_{\pm (\hat{x}+\hat{y})}=0.08$,
 $t^{\alpha\bar{\alpha}}_{\pm (\hat{x} \pm \hat{y})}= -0.12$, 
$t^{\alpha\alpha}_{\pm 2\hat{x}} = t^{\alpha\alpha}_{\pm 2\hat{y}}=0.25$. 
The chemical potential $\mu$ is adjusted to give a fixed filling factor.

The local electronic correlations include the on-site Hubbard repulsion of electrons and Hund's rule coupling of spins.
They are described by the term $\mathcal{H}_\text{I}$, which at the mean-field level takes the form
\begin{eqnarray}
\mathcal{H}_\text{I}^\text{MF}=&&U\sum_{\mathbf{i}\alpha\sigma}\langle\hat{n}_{\mathbf{i}\alpha\bar{\sigma}}\rangle\hat{n}_{\mathbf{i}\alpha\sigma}+U^\prime\sum_{\mathbf{i}\alpha\neq\beta\sigma}\langle\hat{n}_{\mathbf{i}\alpha\bar{\sigma}}\rangle\hat{n}_{\mathbf{i}\beta\sigma}\nonumber\\
&&+(U^\prime-J_H)\sum_{\mathbf{i}\alpha\neq\beta\sigma}
\langle\hat{n}_{\mathbf{i}\alpha\sigma}\rangle\hat{n}_{\mathbf{i}\beta\sigma}\;.
\end{eqnarray}
with the on-site Hubbard potential $U$, the inter-orbital Coulomb repulsion
$U^\prime$, and the Hund's rule coupling $J_H$. The orbital rotation
symmetry imposes the constraint $U=U^\prime+2J_H$.
In Eq.~(\ref{EQ:hamil_tot}), the term $\mathcal{H}_\text{pair}$
contains the effective pairing interaction between two electrons on the NNN site. In mean-field theory this can be written as 
\begin{equation}
\mathcal{H}_\text{pair}=\sum_{\mathbf{i}\mathbf{j}\alpha}(\Delta_{\mathbf{i}\mathbf{j}}^\alpha
d^\dagger_{\mathbf{i}\alpha\uparrow}d^\dagger_{\mathbf{j}\alpha\downarrow}+\text{H.c.})
\delta_{\mathbf{i}\pm\hat{x}\pm\hat{y},\mathbf{j}}\;.
\end{equation}
As has been widely discussed in the literature, this NNN-pairing interaction 
ultimately leads to the proposed $s_\pm$-wave symmetry of iron pnictides.~\cite{mazin08,KKuroki:2008,wang09,YBang:2009,IIMazin:2010}
Finally, the last term $\mathcal{H}_\text{imp}$  in Eq.~(\ref{EQ:hamil_tot}) describes
the scattering potential due to the randomly distributed impurities.
We model the disorder term by a local intra-orbital scattering potential
\begin{eqnarray}\label{H_imp}
\mathcal{H}_\text{imp}&=&\sum_{\mathbf{I}\alpha\sigma}    \{ Wd^\dagger_{\mathbf{I}\alpha\sigma}d_{\mathbf{I}\alpha\sigma} + \delta t 
[d^\dagger_{\mathbf{I}+(-)\hat{x}\alpha\sigma}d_{\mathbf{I}+(-)\hat{x} \pm \hat{y}\bar{\alpha}\sigma}  \nonumber \\
&&+d^\dagger_{\mathbf{I}+(-)\hat{y}\alpha\sigma}d_{\mathbf{I}+(-)\hat{y} \pm \hat{x}\bar{\alpha}\sigma} 
+\text{H.c.}] \}
\; .
\end{eqnarray}
Here the impurity means that an Fe atom at lattice site $\mathbf{I}$ is substituted by Zn atom. Therefore, the on-site energy of the impurity atom is
changed and acts as a nonmagnetic potential scattering center which scrambles the crystal momentum.
As represented by the first term on the rhs of Eq..~(\ref{H_imp}), we consider only intra-orbital scattering. This simplification is justified by our numerical calculations, which show that inter-orbital scattering processes are irrelevant in the unitary limit.
In Eq.~(\ref{H_imp}) we explicitly consider the difference in covalent radii of the Zn atom compared to the Fe atom,
which is captured by the second term proportional to $\delta t$.
Therefore, the substitution introduces an additional change in the hopping parameters among the nearest-neighbor Fe sites of the impurity site.
In the case of the Zn substitution, in addition to electron doping which can be tuned by the chemical potential, the induced local impurity potential is expected to be much stronger than for other transition metals like Co and Ni.
Note that when  the impurity potential on the Zn site is very large, the effect caused by a small change in the Fe-Zn hopping integrals is negligible.
Hence, the surrounding Fe-Fe bond disorder is the second most important term next to the strength of the impurity potential.

We then diagonalize the mean-field Hamiltonian $\mathcal{H}$ of Eq.~(\ref{EQ:hamil_tot}) by
solving the BdG equation self-consistently:
\begin{equation}
\sum_{\mathbf{j}\beta} \left(
\begin{array}{cc}
{\mathcal H}_{\mathbf{i}\mathbf{j}\uparrow}^{\alpha\beta} & \Delta_{\mathbf{i}\mathbf{j}}^\alpha\delta_{\alpha,\beta} \\
\Delta_{\mathbf{j}\mathbf{i}}^{\alpha *}\delta_{\alpha,\beta} & -\mathcal{H}_{\mathbf{j}\mathbf{i}\downarrow}^{\beta\alpha}
\end{array}
\right)
\left(
\begin{array}{c}
u_{\mathbf{j}\beta}^{n} \\
v_{\mathbf{j}\beta}^{n}
\end{array}
\right)
=E_{n} \left(
\begin{array}{c}
u_{\mathbf{i}\alpha}^{n} \\ v_{\mathbf{i}\alpha}^{n}
\end{array}
\right)\;,
\end{equation}
where
$\mathcal{H}_{\mathbf{i}\mathbf{j}\sigma}^{\alpha\beta}=
\tilde{t}^{\alpha\beta}_{\mathbf{i}\mathbf{j}}+(U\langle
\hat{n}_{\mathbf{i}\alpha\bar{\sigma}}\rangle
+U^\prime\sum_{\gamma\neq\alpha\sigma^\prime}\langle\hat{n}_{\mathbf{i}\gamma\sigma^\prime}\rangle
-J_H\sum_{\gamma\neq\alpha}\langle\hat{n}_{\mathbf{i}\gamma\sigma}\rangle
+W\delta_{\mathbf{I},\mathbf{i}}-\mu)\delta_{\alpha,\beta}\delta_{\mathbf{i},\mathbf{j}}$
is the single-particle Hamiltonian and $\tilde{t}$ includes the effect of the local change in the hopping parameter between Fe sites neighboring he impurities,
$\langle\hat{n}_{\mathbf{i}\alpha\uparrow}\rangle=\sum_n|u^n_{\mathbf{i}\alpha}|^2f(E_n)$,
$\langle\hat{n}_{\mathbf{i}\alpha\downarrow}\rangle=\sum_n|v^n_{\mathbf{i}\alpha}|^2[1-f(E_n)]$,
and
$\Delta^\alpha_{\mathbf{i}\mathbf{j}}=(V/2)\sum_n\{u^n_{\mathbf{i}\alpha}v^{n*}_{\mathbf{j}\alpha}[1-f(E_n)]
-v^{n*}_{\mathbf{i}\alpha}u^n_{\mathbf{j}\alpha}f(E_n)\}\delta_{\mathbf{i}\pm\hat{x}\pm\hat{y},\mathbf{j}}$.
Here $V$ is the pairing strength and $f(E)$ is the Fermi-Dirac
distribution function. The local superconducting order parameter 
and charge density at site $\mathbf{i}$ are defined as
\begin{subequations}
\begin{eqnarray}
 \Delta_{\mathbf{i}}&=&\frac{1}{4}\sum_{\mathbf{j}\alpha}
 \Delta^\alpha_{\mathbf{i}\mathbf{j}}\delta_{\mathbf{i}\pm\hat{x}\pm\hat{y},\mathbf{j}}\;,\\
 n_\mathbf{i}&=&\sum_{\alpha\sigma} \langle n_{\mathbf{i}\alpha \sigma}\rangle \;,
\end{eqnarray}
\end{subequations}
respectively.
 Throughout this work,  the
numerical calculations are performed on a $28\times 28$ square
lattice with the periodic boundary condition. A $48\times48$
supercell is taken to calculate the density of states. The interaction parameter values are fixed at 
$(U,J_H,V)=(3.2,0.6,1.05)$;~\cite{YYTai:2013} while the electron filling is chosen to be 2.13 in the pristine system, which corresponds to the optimal electron doping regime. 
For the impurity scattering, we fix
the impurity scattering strength $W=-20$  and $\delta t = -0.2$. This value of $W$ is
reasonable given by the Zn core level located $\sim 8$ eV below the Fermi
level and is very close to the strong scattering (unitary)
limit. Note we have checked several values of impurity scattering strength $W$. The results are  insensitive to the precise value for $W < -8$.
This  might also be relevant to the impurity effect from Co and Ni substitution in LaFeAsO.\cite{Nakamura11}
Specifically, for $W=-0.5$ the superconducting transition temperature is hardly affected by impurities up to $n_{imp} = 25.5\%$.
\cite{rotter08, ALeithe-Jasper:2008, NNi:2008}

\section{Single impurity effects}
\label{sec:single}

\begin{figure}[htp]
\includegraphics[width=0.48\textwidth]{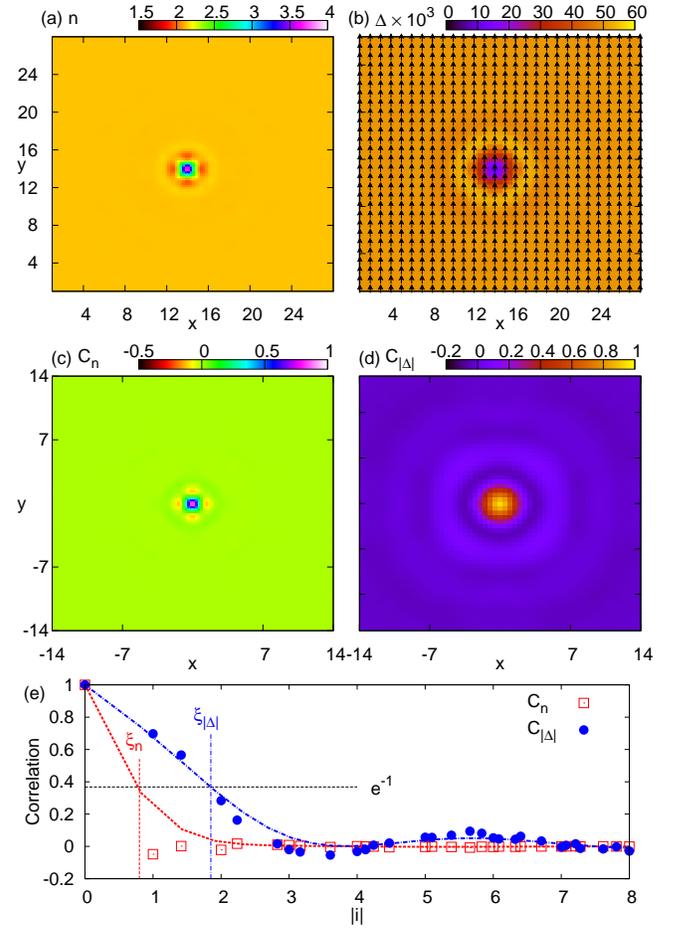}
\caption{\label{fig:si} (Color online) Single impurity effects at zero temperature. The intensity plots of the local charge density (a) and superconducting order parameters (b). The arrows represent the argument of the complex superconducting order. The 2D cross-correlation functions of charge (c) and modulus of order parameter (d). For quantitative analysis the same correlation functions vs.\ distance $|\mathbf{i}|$ in (e).  
}
\end{figure}

Before we proceed with the complex disorder configuration, the single impurity effect on superconducting phase is studied. 
In the absence of impurities, our two-orbital model naturally captures the relation between SDW and SC phases 
and recovers the whole phase diagram with doping evolution. 
Henceforth, we shall restrict our calculations within this set of interaction parameters.
With a single impurity in the unitary limit, we find that impurity scattering induces strong charge inhomogeneity and significantly suppresses the
superconducting order parameter around the impurity site as illustrated in Fig.~\ref{fig:si}.
In particular, we revisit the effects of a single Zn impurity on superconductivity in Ba(Fe$_{1-x-y}$Co$_y$Zn$_x$)$_2$As$_2$  with $y>0.1$, where there is no SDW. 
Note that localization of electrons on the impurity site is taken into account through the modified hopping coefficients of surrounding Fe atoms as presented in Eq.~(\ref{H_imp}).

To gain deeper physical insights into scattering effects around a single impurity,
we consider the 2D spatial cross-correlation functions of the superconducting order
parameter and charge density defined by
 \begin{equation}
 C_X(\mathbf{i})=
 \frac{ \sum_\mathbf{j}  [(X(\mathbf{i}+\mathbf{j})-\langle X\rangle)(X(\mathbf{j})-\langle X \rangle)]}{\sum_{\mathbf{j}} [X(\mathbf{j}) - \langle X\rangle]^2}
 \label{eq:correlation}
 \end{equation}
where $X=n$ and $|\Delta|$, and  the mean $\langle X \rangle=(1/N)\sum_\mathbf{j} X(\mathbf{j})$ with $N$ the number of lattice sites.
The cross-correlation function is normalized to give $-1 \leq C_X \leq 1$.
The results for the 2D cross-correlation functions are plotted in
Fig.~\ref{fig:si}(c) through ~\ref{fig:si}(e), where the fourfold symmetry and rapid screening over a few lattice sites becomes obvious.
A quantitative analysis is possible when plotting $C_X$ as a function of distance from the impurity site. In Fig.~\ref{fig:si}(e) we define a typical spatial correlation length
$\xi_X$ by measuring the impurity-induced fluctuations of $X$ as the distance where $C_X$ drops from unity to $1/e$. 
It is straightforward to read off  from Fig.~\ref{fig:si}(e) that the additional local charge on Zn is well-screened within a lattice distance, $\xi_n \sim 1$. Indeed, it is over screened, resulting in Friedel-type oscillations, which are clearly visible in the correlation function.
Such a short screening length is mainly due to the strong local Coulomb repulsion $U$, which acts on the charge sector.
In contrast, the superconducting correlation function has a more profound oscillating tail with a short coherence length $\xi_\Delta \sim 2 $.
Based on these quite short correlation lengths, we expect that the Zn-doped Ba-112 iron-based superconductor  will be a good candidate for the Swiss cheese model,~\cite{BNachumi:1996}
where the holes of the Swiss cheese correspond to the holes punched into the superconducting texture by the Zn impurity, while the effect on the bulk value of the superconducting order parameter is almost negligible after a few lattice sites away from the defect.
Hence we anticipate that the $s_\pm$-pairing gap is easily destabilized by strong impurity scattering similar to the high-$T_c$ cuprate,\cite{BNachumi:1996} Sr$_2$RuO$_4$,\cite{Mackenzie1998} UPt$_3$,\cite{Kycia1998}
and PuCoGa$_5$ superconductors.\cite{Jutier2008}
Indeed, this result is in agreement with available experimental
observations in Ba(Fe$_{1-x-y}$Zn$_x$Co$_y$)$_2$As$_2$, which has a relatively low
Neel temperature $T_N\sim$~135 K and no trace of superconductivity for
 $y=0$ and $x=0.08,  0.25$.~\cite{ideta13a, ideta13b}
On the other hand, when Co doping induces superconductivity, doping by several percent of Zn rapidly suppresses it.

\section{Disorder effects in the superconducting order parameter}
\label{sec:superconductivity}

\begin{figure}[htp]
\includegraphics[width=0.48\textwidth]{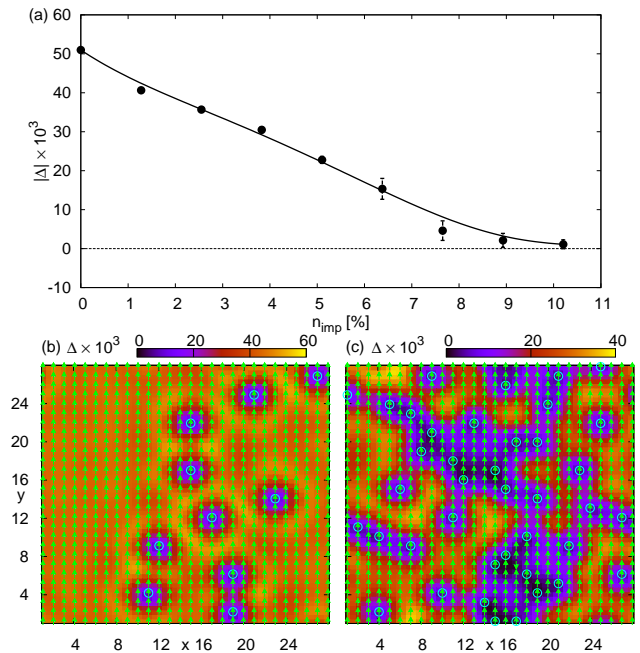}
\caption{\label{fig:sc} (Color online)  Disorder effects at zero temperature. (a) The averaged superconducting order parameter $\langle\Delta \rangle=|\Delta|e^{i\phi}$ as a funtion of $n_\text{imp}$. 
 The data is averaged over 5 disorder configurations.  
The error bars on the data points represents the statistical deviation.
The solid line is fitted to guide the eye. 
Intensity plot of local superconducting order parameters for a typical impurity configuration with concentration $n_\text{imp}=1.28\%$ (b) and $n_\text{imp}=5.10\%$ (c). The arrows and open green circles represent the argument of complex superconducting order parameter and the impurity position respectively.}
\end{figure}

We next turn to the question of how superconductivity is affected by
increasing the impurity concentration. For this purpose, the
evolution of the disorder-configuration-averaged superconducting
order parameter $\langle\Delta \rangle$ (at  zero temperature)
as a function of the impurity concentration $n_\text{imp}$ is
considered. 

As for the case of the single impurity study, we focus on the compound Ba(Fe$_{1-x-y}$Zn$_x$Co$_y$)$_2$As$_2$, with $y>0.1$ when the SDW phase is suppressed. Again we are primarily interested in the local suppression of the superconductivity due to Zn substitution and the combined effects of charge localization and strong impurity scattering. For that purpose we make the following simplifications: (1) doping with Co adds mainly charge to the itinerant electrons that is captured by a shift of the chemical potential, and (2) scattering is in the weak limit compared to Zn. Hence the local scattering effect of Co impurities is neglected.
The results of the suppression of the lattice-averaged order parameter are plotted in Fig.~\ref{fig:sc}(a). We find
that the averaged modulus of the order parameter $\langle|\Delta| \rangle$ decays
nearly linearly with increasing impurity concentration and eventually
vanishes at a critical concentration of $n_\text{imp}\approx 10-11\%$. 
Considering the superconducting transition temperature $T_\text{c}$ is usually over estimated at the mean-field level (more on this latter in Sec.~\ref{sec:dos}),
our results are in good agreement with recent measurements in Ba(Fe$_{1-x-y}$Zn$_x$Co$_y$)$_2$As$_2$.~\cite{li12a}

To provide an intuitive picture of disorder effects in highly disordered
superconductors with increasing impurity concentration, we present a
study of the evolution of the local superconducting order parameter for
two particular realizations of disorder configurations. The spatially
resolved order parameter $\Delta_\mathbf{i}$  is shown in colormaps in
Figs.~\ref{fig:sc}(b) and (c) for $n_\text{imp}=1.28\%$ and
$5.10\%$, respectively. The images reveal that the order parameter is locally suppressed at
the impurity sites, and the impurities behave individually when the
impurity concentration is small as shown in Fig.~\ref{fig:sc}(b).
Of great interest is that the interference of the local order
parameter at each impurity site develops gradually with increasing
impurity concentration $n_\text{imp}$, as one can clearly observe
from Fig.~\ref{fig:sc}(c), where islands form. The crude estimation on the threshold
length of interference is given by $\xi_\Delta$ as illustrated in the previous
Fig.~\ref{fig:si}(e). Also a considerable portion of sites has vanishing
order parameter amplitude in the highly disordered limit. These correlated sites
form islands and break the system into several superconducting
puddles as illustrated in Fig.~\ref{fig:sc}(c). 
Consequently, the local
order parameter becomes highly inhomogeneous in Fig.~\ref{fig:sc}(a).
 We propose, as in the case of high-temperature cuprate superconductors,~\cite{AVBalatsky:2006}  that the novel
electronic inhomogeneity should also be detected by measuring the local
density of states using the atomic resolution scanning tunneling
microscopy.

\begin{figure} [htp]
\includegraphics[width=0.45\textwidth]{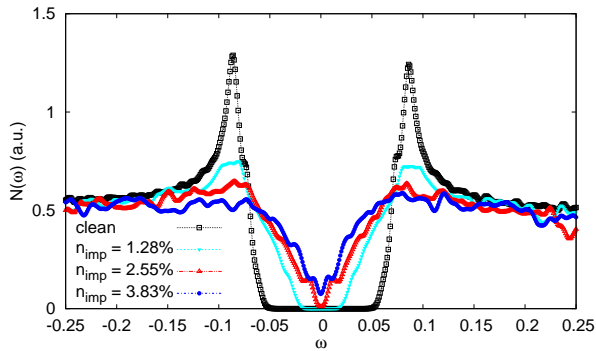}
\caption{\label{fig:dos} (Color online) Total density of states for various sets of impurity concentrations $n_\text{imp}=1.28, 2.55, 3.83\%$ at very low temperature $T=0.002 \ll T_c$. }
\end{figure}

\section{Total density of states and superfluid density}
\label{sec:dos}

To gain further insight into the disorder effects in the unitary
limit of highly disordered superconductors, we calculate several observables such as the total density
of states (DOS), the superfluid density, and the magnetic penetration depth $\lambda$. 
The site-averaged DOS at finite temperature is defined by
\begin{equation}
N(\omega)=-\frac{1}{N}\sum_{\mathbf{i}\alpha
n}[|u_{\mathbf{i}\alpha}^n|^2f^\prime(\omega-E_n)+|v_{\mathbf{i}\alpha}^n|^2f^\prime(\omega+E_n)],
\end{equation}
where $f^\prime(E)$ is the derivative of the Fermi-Dirac
distribution function with respect to the Fermi energy. For better
visualization in Fig.~\ref{fig:dos}, the DOS is calculated at finite
temperature $T=0.002 \ll T_c$. Note that in the pristine system 
two BCS coherence peaks are exhibited at the energies $\omega=\pm0.1$, which corresponds to the single particle excitation gap.
With increasing impurity concentration, the coherence peaks are gradually suppressed. Eventually above 
$n_\text{imp}=3.83\%$, the DOS is filled in and gapless superconductivity emerges.

In experiments, the magnitude of the superconducting transition
temperature $T_c$ is usually less sensitive to defects since it is
related to the spatial average of the order parameter, which is a local correlation function. On the other side, the
magnitude of the penetration depth $\lambda$
measures the stiffness of the superconducting phase coherence in the superconductor, which is a nonlocal response function.
Therefore, this quantity can provide deep insight into the nature of the superconducting pairing symmetry through its temperature dependence and residual value
because these are extremely sensitive to defects. 
So far,  measurements of the magnetic penetration penetration 
depth in Fe-SCs have given controversial results. 
For example, in 122-type iron pnictides, the  superfluid density exhibits an exponential temperature behavior in the cleanest hole-doped compounds,
Ba$_{1-x}$K$_x$Fe$_2$As$_2$,\cite{KHashimoto:2009} 
while a power-law behavior is seen in Ba(Fe$_{1-x}$Co$_x$)$_2$As$_2$.\cite{gordon10,RTGordon:2009a,RTGordon:2009b,HKim10,JYong:2011,AABarannik:2011,TJWilliam:2009}
Very recently, two of us and co-workers studied the temperature dependence of the superfluid density of clean 122-type iron pnictides  at various electron-doping levels and  found that the low-temperature power-law dependence of the deviation
$\Delta \lambda(T)=\lambda(T)-\lambda(0)$ varies with an exponent greater than $3$.~\cite{HHuang:2012} 

\begin{figure}[htp]
\includegraphics[width=0.48\textwidth]{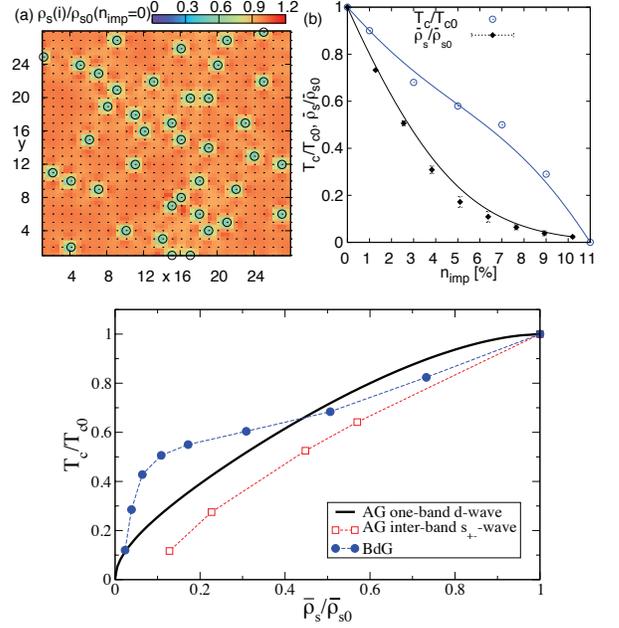}
\caption{\label{fig:sf} (Color online) (a) The intensity of
local superfluid density $\rho_s(\mathbf{i})$ with $n_\text{imp}=3.83\%$ at zero temperature. The
open black circles indicate the impurity locations. (b) The
zero-temperature bulk superfluid density $\bar\rho_s$ and superconducting transition temperature
$T_\text{c}$ as a function of impurity concentration
$n_\text{imp}$.  The data is averaged over five randomly distributed impurity configurations. (c) The Uemura plot of the superfluid density in short-coherence superconductors.  The variables $T_{c0}$ and $\bar\rho_{s0}$ are obtained from a pristine system. For comparison, we also plot results of the one-band AG calculations for $d$-wave pairing symmetry~\cite{TDas:2011} and the two-band AG calculations for the $s_{\pm}$-wave symmetry.~\cite{ABVorontsov:2009} 
 }
\end{figure}

In our multiorbital lattice BdG calculations, we follow the standard linear response approach of Refs.~\onlinecite{scalapino92} and \onlinecite{scalapino93} to investigate disorder effects on the superfluid density.  
In the presence of a weak vector potential
$\mathcal{A}_\eta(\mathbf{r},t)$ along the direction $\eta$, the
hopping term is modified by the Peierls phase factors $e^{i\int
\mathcal{A}_\eta(\mathbf{r},t)d\mathbf{r}}$ (We set $e=\hbar=c=1$). 
Hence the change in the  tight-binding Hamiltonian in the Meissner state is
\begin{eqnarray}
&&\mathcal{H}^\prime_0=\sum_{\mathbf{i}\mathbf{\delta}
\alpha\beta\sigma}t^{\alpha\beta}_{\mathbf{i}\mathbf{i}+\mathbf{\delta}}
d^\dagger_{\mathbf{i}\alpha\sigma}d_{\mathbf{i}+\mathbf{\delta}} 
\nonumber\\
&&
\times [-i\mathcal{A}_\eta(\mathbf{i},t)\mathbf{\delta}_\eta
-\frac{1}{2}(\mathcal{A}_\eta(\mathbf{i},t)\mathbf{\delta}_\eta)^2] 
+\mathcal{O}(\mathcal{A}_\eta^3)
\end{eqnarray}
where $\mathbf{\delta}_\eta$ projects $\mathbf{\delta}$ onto the
direction $\eta$ in units of the lattice constant. The charge
current density operator consists of the usual paramagnetic and
diamagnetic parts,
 \begin{equation}
 \hat{j}_\eta(\mathbf{i},t)=-\frac{\partial \mathcal{H}^\prime_0}{\partial \mathcal{A}_\eta(\mathbf{i},t)}=\hat{j}^P_\eta(\mathbf{i},t)+\hat{j}^D_\eta(\mathbf{i},t),
 \end{equation}
 with
 \begin{equation}
 \{\hat{j}^P_\eta(\mathbf{i},t),\hat{j}^D_\eta(\mathbf{i},t)\}=
 \sum_{\mathbf{\delta}\alpha\beta\sigma}t^{\alpha\beta}_{\mathbf{i}
 \mathbf{i}+\mathbf{\delta}}d^\dagger_{\mathbf{i}\mathbf{\delta}
 \sigma}d_{\mathbf{i}+\mathbf{\delta}\beta\sigma}\mathbf{\delta}_\eta\{i,\mathcal{A}_\eta \mathbf{\delta}_\eta\}. 
 \end{equation}
 In the interaction representation, the kernel function $K$ of the charge current satisfies
  \begin{equation}
  \langle \hat{j}_\eta(\mathbf{i},t)\rangle =
  -\sum_{\mathbf{i}^\prime}\int dt^\prime K(\mathbf{i},\mathbf{i}^\prime, t-t^\prime)\mathcal{A}_\eta(\mathbf{i}^\prime,t^\prime)
  \end{equation}
 to leading order in the vector potential $\mathcal{A}_\eta$, where, the static kernel at $\omega=0$ is expressed by
\begin{eqnarray}
  K(\mathbf{i},\mathbf{i}^\prime,\omega=0)=-\sum_{nm}
  \Gamma^{nm}_\mathbf{i}\Gamma^{mn}_{\mathbf{i}^\prime}
  \frac{f(E_m)-f(E_n)}{E_m-E_n}\nonumber\\
  -\sum_{\mathbf{\delta}\alpha\beta
  n}t^{\alpha\beta}_{\mathbf{i}\mathbf{i}+\mathbf{\delta}}
  [u_{\mathbf{i}\alpha}^{n*}u_{\mathbf{i}+\mathbf{\delta}\beta}^n
  f(E_n)+v_{\mathbf{i}\alpha}^nv_{\mathbf{i}+\mathbf{\delta}\beta}^{n*}
  f(-E_n)]\mathbf{\delta}_\eta^2\delta_{\mathbf{i},\mathbf{i}^\prime}.
  \nonumber\\
\end{eqnarray}
Here, the auxiliary functions are
$\Gamma^{nm}_\mathbf{i}=\sum_{\mathbf{\delta\alpha\beta}}
t^{\alpha\beta}_{\mathbf{i}\mathbf{i}+\mathbf{\delta}}
(u_{\mathbf{i}\alpha}^{n*}u_{\mathbf{i}+\mathbf{\delta}\beta}^m
-v_{\mathbf{i}\alpha}^mv_{\mathbf{i}+\mathbf{\delta}\beta}^{n*})\mathbf{\delta}_\eta$.
Fourier transform with respect to the individual coordinates
$\mathbf{i}$ and $\mathbf{i}^\prime$ then defines the spatially
averaged kernel function
$\bar K(\mathbf{q},\omega=0)=(1/N)\sum_{\mathbf{i},\mathbf{i}^{\prime}}e^{-i\mathbf{q}
\cdot(\mathbf{i}-\mathbf{i}^{\prime})}K(\mathbf{i},\mathbf{i}^\prime,\omega=0)$, 
which gives the bulk superfluid density $\bar{\rho}_s= \bar K(\mathbf{q}\rightarrow 0, \omega=0)$.
We also define the local superfluid density as
\begin{equation}
\rho_s(\mathbf{i})=K(\mathbf{i},\mathbf{i};\omega=0)
\end{equation} 
to investigate the local suppression of the superfluid density.
As shown in  Fig.~\ref{fig:sf}(a), we find that the local superfluid
density is dramatically suppressed at impurity sites. As
illustrated in Fig.~\ref{fig:sf}(b), the bulk superfluid density $\bar\rho_s$
decreases drastically to zero, much faster than $T_c$, as expected with increasing impurity concentration
$n_\text{imp}$ in the Swiss cheese scenario for a short coherence superconductor.
This different rate of suppression is further corroborated by the Uemura plot as shown in Fig.~\ref{fig:sf}(c), suggesting the break-down of the AG theory. 
We note that the loss of phase coherence is related to the
vanishingly small superfluid density near the critical
concentration of impurities,\cite{bouadim11} implying the importance of spatial disorder induced fluctuations.\cite{ghosal01,benfatto04} 
As manifested in
Fig.~\ref{fig:sc}(c), a Bose system consisting of localized Cooper
pairs is gradually formed in the highly disordered limit due to the loss of
phase coherence between the superconducting puddles. Unfortunately, 
the physically interesting region, where the superfluid density is vanishing small, 
is not captured within the BdG framework due to the neglect of phase fluctuations. 
In the present mean-field theory the
phases of the order parameter at different sites are completely
aligned with the ground state as shown in Fig.~\ref{fig:sc}(b) and
(c). For details on the consequences of quantum phase fluctuations on the order
parameter in the
inhomogeneous BdG state using a quantum XY model see Ref.~\onlinecite{Ramakrishnan89}.

\begin{figure}[htp]
\includegraphics[width=0.45\textwidth]{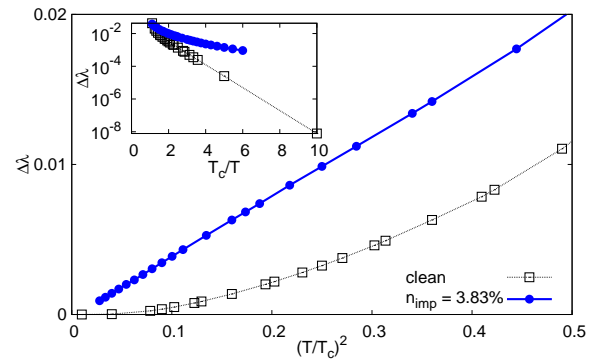}
\caption{(Color online) Temperature dependence of the deviation of the magnetic penetration depth $\Delta\lambda$
in the clean limit (open black square) and for a dirty system with impurity concentration $n_\text{imp}=3.83\%$ (solid blue circle). 
The data is averaged over five randomly selected disorder configurations.
Inset: Replotted data to emphasize the exponential low-temperature behavior of the clean system.}
\label{fig:London} 
\end{figure}

Finally, we also calculated the temperature dependence of the deviation $\Delta \lambda(T)$ of the magnetic penetration depth in the presence of disorder, which is related to the bulk superfluid density $\lambda^2 \propto 1/\bar\rho_s$.
In the clean limit, $\Delta \lambda(T)$ is expected to vary exponentially 
at low temperatures due to a gapped DOS, as shown in Fig.~\ref{fig:London}.
The exponential decay is consistent with a fully gapped pairing state.
At the impurity concentration $n_\text{imp}=3.83\%$, 
the temperature dependence of $\Delta \lambda(T)$ shows a $T^2$ power law. Hence we expect that for intermediate impurity concentrations the temperature behavior will resemble that of a power law with exponent greater than two.
Interestingly, the $T^2$ variation of $\Delta \lambda(T)$ is observed experimentally in Ba$_{1-x}$K$_x$Fe$_2$As$_2$,~\cite{Hashimoto09,CMartin:2009}  and Ca$_{0.5}$Na$_{0.5}$Fe$_2$As$_2$ single crystals,~\cite{JKim:2012} possibly due to the doping-induced disorder.  
Our calculation showcases that the temperature dependence of the penetration depth in an $s_{\pm}$-wave pairing superconductor can be very sensitive to the impurity scattering. Depending on the impurity concentration, it may enable us to explain various power-law behaviors in Fe-SCs.~\cite{gordon10,RTGordon:2009a,RTGordon:2009b,HKim10,JYong:2011,AABarannik:2011,TJWilliam:2009}

\section{summary}
\label{sec:summary} To summarize, by solving the
lattice BdG equations self-consistently, we have studied disorder effects on superconducting and transport properties  of disordered superconductors with $s_{\pm}$
pairing symmetry. In the unitary limit, the impurity
scattering strength is so large that the potential scattering
term cannot be treated as a perturbation in the framework of pair breaking by Abrikosov and Gorkov. The detailed numerical
calculations demonstrate that a single nonmagnetic impurity
can depress  superconductivity significantly at the local scale.
 With increasing impurity concentration the
impurity scattering potential induces a spatial redistribution of
the amplitudes of local Cooper pairs in the form of superconducting puddles, giving rise to 
significant spatially electronic inhomogeneity. 
Calculations of the local
density of states, the superfluid density, as well as the magnetic
penetration depth further reinforce this picture, demonstrating
again that the superconducting phase is not stable against 
strong impurity scattering as expected in the Swiss cheese scenario.

Our results  shed new light on the understanding of recent experiments in Co- and Zn-substituted BaFe$_2$As$_2$ samples.\cite{li12a} 
In these samples the superconductivity is completely suppressed when the concentration of Zn impurities $n_{\text{imp}}$ is above
$8\%$.
The available angle-resolved photoemission spectrocopy experiments
indicate that the substitution by Zn atoms not only provides additional
electrons into the Fe lattice, but also creates strong local scattering
potentials because the Zn-$3d$ orbitals are well-below the Fermi
level.\cite{ideta13a, ideta13b} 
All these observations are consistent with our numerical results. 
Furthermore, our calculations show that superconductivity is hardly affected by weak intraorbital scattering 
with scattering potential $W=-0.5$ (corresponding to Co and Ni) or by interorbital scattering in the unitary limit.
We anticipate that the emergent electronic inhomogeneity in the strong scattering limit, due to local screening effects, will be probed in future scanning tunneling microscope and scanning Meissner force microscope experiments.~\cite{JHXu:1995}

\section*{Acknowledgments}
We thank Y. Chen, S. Zhou, Y.K. Li, I. Vekhter, Y. Gao, and H. Huang for very useful discussions. This work was supported in part by the NSF of China under Grant No.11274084,
the NSF of Zhejiang Province under Grant No. Z6110033, and the 973 Project of the MOST under Grant No. 2010CB923000 (H.C. \& J.D.),  
by the Robert A. Welch Foundation under Grant No. E-1146 (Y.-Y.T. \& C.S.T.), and by LANL LDRD Program (M.J.G. \& Y.-Y.T.).  This work was, in part, supported by the Center for Integrated Nanotechnologies,
a U.S. DOE Office of Basic Energy Sciences user facility. 
M.J.G. also thanks  the Aspen Center for Physics for its hospitality, which is supported by the NSF under Grant No. PHYS-1066293.
We are grateful for a computation allocation at the High Performance Computing
Clusters at the Institute for Fusion Theory and Simulation of Zhejiang University.

\appendix
\section{Unfolding transformation of the tight-binding model}
\label{Appendix}

Here we derive the unitary transformations for the rotation of orbitals between both Fe sublattices to attain the exact mapping of the 2-by-2 orbital model onto the model of two decoupled two-orbital Hamiltonians.

\begin{SCfigure*}
\includegraphics[scale=0.3,angle=0]{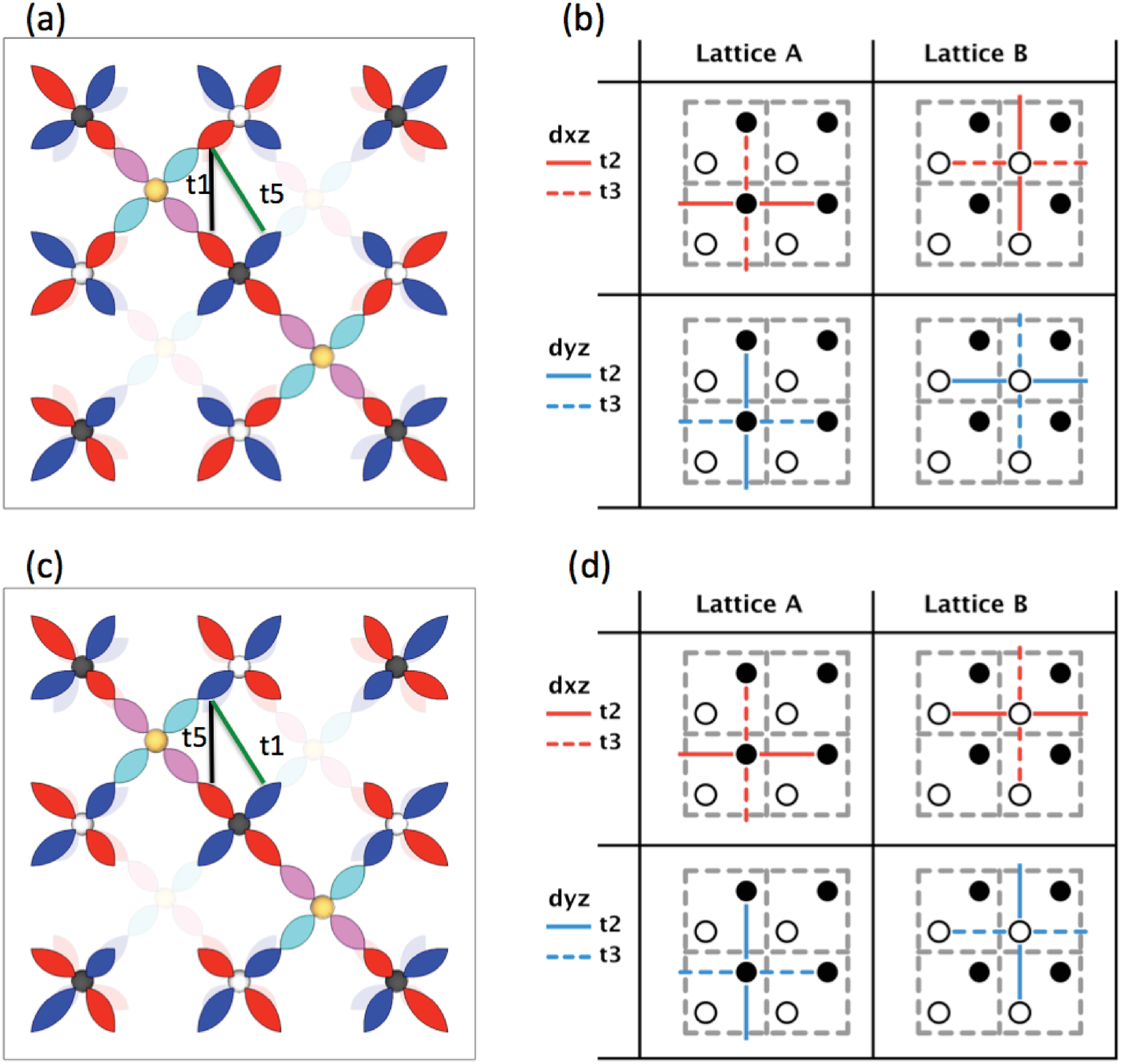}
\caption
{(Color online) Two choices of the basis (a), (b)[(c), (d)] with[without] a $90^\circ$ rotation of the local coordinate system on sublattice $B$. Panels (a), (c) show the $d_{xz}$ and $d_{yz}$ orbital symmetry and the overlap through As-$p_{x/y}$ orbitals: the NNN intra- (inter-) hopping terms $t_1$ ($t_5$) are indicated by the black (green) solid lines. 
Panels (b), (d) illustrate the NNN intraorbital hopping terms for $t_2$($t_3$) in solid (dashed) lines. 
Note that the coordinates of (a) and (c) have a $45^\circ$ rotation from (b) and (d).}
\label{fig:twist}
\end{SCfigure*}

{\it Description of the orbital twist argument.}
As proposed in Ref.~\onlinecite{YYTai:2013}, 
the tight-binding Hamiltonian of Eq.~(\ref{EQ:hamil_tot}) 
in the 2-Fe unit cell Brillouin zone (BZ) is given by 
$\mathcal{H}_0 = \sum_{{\mathbf k}}\psi^\dagger({{\mathbf k}})\, \mathbb{W}_{{\mathbf k}}\, \psi({{\mathbf k}})$ 
 with
\begin{equation}
\begin{aligned}
\label{eq:wk}
\mathbb{W}_{{\mathbf k}} =& 
\left(
\begin{array}{cccc}
 \xi^{H}-\mu& \xi_{12}& \xi_{t}&  \xi_{c} \\
 \xi_{12}& \xi^{V}-\mu& \xi_{c}&  \xi_{t} \\
 \xi_{t}&  \xi_{c}&  \xi^{V}-\mu& \xi_{12}\\
 \xi_{c}&  \xi_{t}&  \xi_{12}& \xi^{H}-\mu\\
\end{array}
\right).
\end{aligned}
\end{equation}
Here the four-component field operator is defined as 
$\psi$=$(d_{A1}$,$d_{A2}$,$d_{B1}$,$d_{B2})^T$
with $A$, $B$ labeling the sublattice and $1$($2$) labeling the orbital $d_{yz}$($d_{xz}$).
The dispersions are given by
$\xi^{H}  = 2 t_2 \cos k_x+2 t_3\cos k_y+4 t_6\cos k_x\cos k_y$,
$\xi^{V}  = 2 t_3 \cos k_x+2 t_2\cos k_y+4 t_6\cos k_x\cos k_y$,
$\xi_{12} = 2 t_4 (\cos k_x+\cos k_y)$,
$\xi_{t } = 4 t_1 \cos \frac{k_x}{2}\cos \frac{k_y}{2}$,
$\xi_{c } = 4 t_5 \cos \frac{k_x}{2}\cos \frac{k_y}{2}$ with 
$t_{1-6}=(-1, 0.08, 1.35, -0.12, 0.09, 0.25)$. 
In Eq.~(\ref{eq:wk}), the $C_4$ symmetry of intra-orbital hopping processes between sublattices $A$ and $B$ is broken.
As we will show below, there is a degree of freedom to write the Hamiltonian by rotating the local coordinate on the sublattice $A$ or $B$.
The above $C_4$ symmerty is recovered by a $90^\circ$ rotation of $d_{xz}$ and $d_{yz}$ orbitals on the sublattice $B$ as illustrated in Fig.~\ref{fig:twist}(a). 
Specifically, we define a new basis under the unitary transformation 
$\phi$ = $(d_{A1}^\prime,d_{A2}^\prime,d_{B1}^\prime,d_{B2}^\prime)^T$ = $U \psi$ with
\begin{equation}
\begin{aligned}
\label{utrans}
U =& 
\left(
\begin{array}{cccc}
 1	&	0	&	0	&	0\\
 0	&	1	&	0	&	0\\
 0	&	0	&	0	&	1\\
 0	&	0	&	1	&	0\\
\end{array}
\right).
\end{aligned}
\end{equation}
Namely, the unitary transformation $U$  flips the orbitals $d_{xz}$ and $d_{yz}$ on the sublattice $B$.
The corresponding Hamiltonian has the form
$\mathcal H_0 = \sum_{\bf k} \phi^\dagger({\bf k})\mathbb{W}_{\bf k}^\prime\phi({\bf k})$ with
\begin{equation}
\begin{aligned}
\label{eq:wk1}
\mathbb{W}_{{\bf k}}^\prime =U\mathbb{W}_\mathbf{k}U^\dagger& 
=\left(
\begin{array}{cccc}
 \xi^{H}-\mu&\xi_{12}		&\xi_{c}	&\xi_{t}		\\
 \xi_{12}	& \xi^{V}-\mu	&\xi_{t}	&\xi_{c}		\\
 \xi_{c}	& \xi_{t}		&\xi^{H}-\mu&\xi_{12}		\\
 \xi_{t}	& \xi_{c}		&\xi_{12}	&\xi^{V}-\mu	\\
\end{array}
\right).
\end{aligned}
\end{equation}

\begin{SCfigure*}  
\includegraphics[scale=0.5,angle=0]{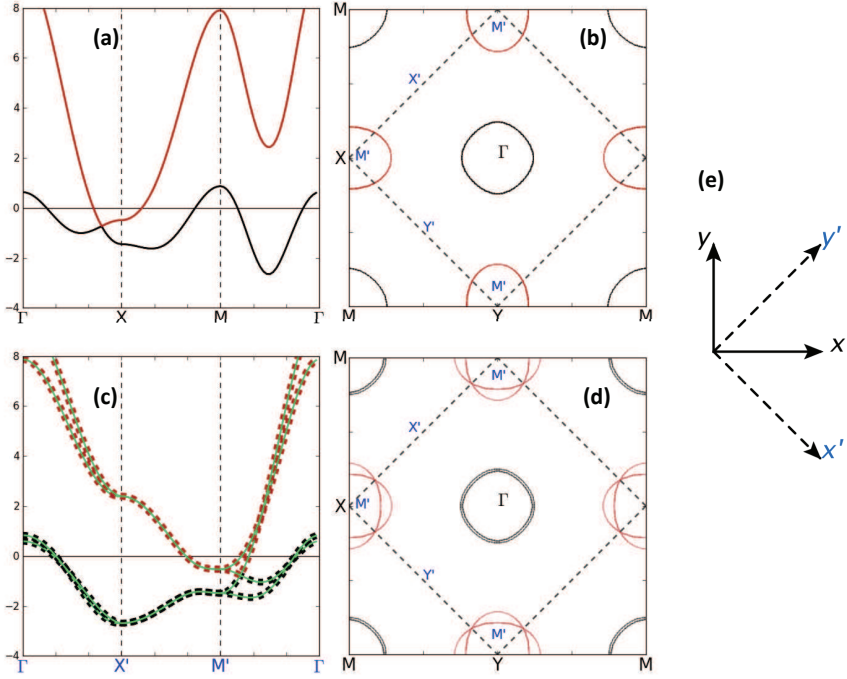}
\caption{(Color online) Calculated band structure (a) and Fermi surface (b) in unfolded (1-Fe per unit cell) BZ.
Here the Fermi energy is shifted to zero for $1/2$ filling.
(c) The folded band structure of the 1-Fe  (green solid line) to  2-Fe per unit cell band structure (dashed line), 
which is identical to that calculated directly from the Hamiltonian before the gauge transformation.
(d) The Fermi surfaces of the 2-Fe band structure. (e)
The transformation between the coordinates in the 1-Fe per unit cell (solid lines) and 2-Fe per unit cell (dashed lines) systems.}
\label{pic:BandFS}
\end{SCfigure*}

{\it Mapping onto the $1$-Fe per unit cell Hamiltonian.} 
Note that $\mathbb{W}_\mathbf{k}^\prime$ in Eq.~(\ref{eq:wk1})
has the same $2\times 2$ block matrix for sublattices $A$ and $B$.
By the symmetry analysis, the entire Hamiltonian can be written in the basis $\phi = (d_1, d_2)^T$ of the $1$-Fe unit cell . The resulting Hamiltonian $\mathcal{H}^0$=$\sum_{\bf k} \phi^\dagger_{\bf k} \mathbb{M}_{\bf k} \phi_{\bf k}$ takes the following form
\begin{equation}
\begin{aligned}
\mathbb{M}_{{\bf k}} =& 
\left(
\begin{array}{cc}
 \xi_{1}-\mu	& \xi_{12}		\\
 \xi_{21}		& \xi_{2}-\mu	\\
\end{array}
\right),
\end{aligned}
\end{equation}
where $\xi_1$ = $E_x+E_t$, $\xi_2$ = $E_y+E_t$ and $\xi_{12}$ = $\xi_{21}$ = $E_c$.
Each component is defined as
\begin{equation}
\begin{aligned}
	E_t 	&= 2t_1[\cos k_x+\cos k_y]+2t_6[\cos 2k_x+\cos 2k_y],\\
	E_x 	&= 2(t_2+t_3)\cos k_x\cos k_y +2(t_2-t_3)\sin k_x\sin k_y,\\
	E_y 	&= 2(t_2+t_3)\cos k_x\cos k_y -2(t_2-t_3)\sin k_x\sin k_y,\\
	E_c		&= 2t_5[\cos k_x+\cos k_y]+4t_4\cos k_x\cos k_y,\\
\end{aligned}
\end{equation}
with a new set of hopping parameters  $t_{1-6}=(0.09, 0.08, 1.35, -0.12, -1, 0.25)$.
Figure~\ref{pic:BandFS}(a) and (b) shows the band structure and Fermi surface with half electron filling in the BZ  corresponding to 1-Fe per unit cell and the Dirac dispersions can be observed around $X$ and $Y$ points.
The comparison between the 1-Fe band structure and 2-Fe band structure is shown in Fig.~\ref{pic:BandFS}(c). 
The 1-Fe band structure can be nicely folded onto the 2-Fe band structure. 
The folded Fermi surfaces are also presented in Fig.~\ref{pic:BandFS}(d).
The corresponding band dispersions in the reduced BZ are given by the block-structured matrix
\begin{equation}
\begin{aligned}
\mathbb{W}_{{\bf k}}^{\prime\prime} =& 
\left(
\begin{array}{cc}
 \mathbb{M}_{\bf k}	& 		0	\\
 0		& \mathbb{M}_{\bf k+Q}	\\
\end{array}
\right),
\end{aligned}
\end{equation}
with the folding vector $\mathbf{Q}=(\pi,\pi)$.

We next prove that the $\mathbb{W}_{\bf k}^{\prime\prime}$ is just a gauge transform from $\mathbb{W}_{\bf k}^\prime$.
The explicit form of band dispersions in $2$-Fe unit cell is given by
\begin{equation}
\begin{aligned}
	\{\xi^H,\xi^V\}	=& \{t_2,t_3\}\,[e^{i\mathbf{k}\cdot \hat x}+e^{-i\mathbf{k}\cdot \hat x}]+ \{t_3,t_2\}\,[e^{i\mathbf{k}\cdot \hat y}+e^{-i\mathbf{k}\cdot \hat y}]\\
		+t_6&\,[e^{i\mathbf{k}\cdot (\hat x+\hat y)}+ e^{-i\mathbf{k}\cdot (\hat x+\hat y)}+ e^{i\mathbf{k}\cdot (\hat x-\hat y)}+ e^{-i\mathbf{k}\cdot (\hat x-\hat y)}],\\
	\{\xi_{t},\xi_{c}\} =& \{t_1,t_5\}\,[e^{i\mathbf{k}\cdot (\hat x+\hat y)/2}+e^{-i\mathbf{k}\cdot (\hat x+\hat y)/2}\\
						+&e^{i\mathbf{k}\cdot (\hat x-\hat y)/2}+e^{-i\mathbf{k}\cdot (\hat x-\hat y)/2}],\\
	\xi_{12}=& t_4\,[e^{i\mathbf{k}\cdot \hat x}+e^{-i\mathbf{k}\cdot \hat x}+e^{i\mathbf{k}\cdot \hat y}+e^{-i\mathbf{k}\cdot \hat y}].\\
\end{aligned}
\end{equation}
By the help of the re-definition of ($\hat x, \hat y$) $\rightarrow$ ($\hat x+\hat y, \hat x-\hat y$),
The band dispersions written in $1$-Fe unit cell basis have the from 
\begin{equation}
\begin{aligned}
	\{\xi^H,\xi^V\}	=& \{t_2,t_3\}\,[e^{i\mathbf{k}\cdot (\hat x+\hat y)}+e^{-i\mathbf{k}\cdot (\hat x+\hat y)}]\\
			+&\{t_3,t_2\}\,[e^{i\mathbf{k}\cdot (\hat x-\hat y)}+e^{-i\mathbf{k}\cdot (\hat x-\hat y)}],\\
			+& t_6\,[e^{i\mathbf{k}\cdot (2\hat x)}+ e^{-i\mathbf{k}\cdot (2\hat x)}+ e^{i\mathbf{k}\cdot (2\hat y)}+ e^{-i\mathbf{k}\cdot (2\hat y)}],\\
	\{\xi_{t},\xi_{c}\} =& \{t_1,t_5\}\,[e^{i\mathbf{k}\cdot \hat x}+e^{-i\mathbf{k}\cdot \hat x}+e^{i\mathbf{k}\cdot \hat y}+e^{-i\mathbf{k}\cdot \hat y}],\\
	\xi_{12}=& t_4\,[e^{i\mathbf{k}\cdot (\hat x+\hat y)}+e^{-i\mathbf{k}\cdot (\hat x+\hat y)}+e^{i\mathbf{k}\cdot (\hat x-\hat y)}+e^{-i\mathbf{k}\cdot (\hat x-\hat y)}],\\
\label{eq:newele}
\end{aligned}
\end{equation}
Then we shall consider $\mathbb{W}_{\bf k}^\prime$ with the new elements of Eq.~(\ref{eq:newele}) 
and $\mathbf{k}$ running over the BZ corresponding to 1-Fe per unit cell. Here, we rewrite $\mathbb{W}_{\bf k}^\prime$ in the block matrix form
\begin{equation}
\begin{aligned}
\label{wk1}
\mathbb{W}_{\bf k}^\prime =& 
\left(
\begin{array}{cc}
 \mathbb{A} 	&	\mathbb{B}	\\
 \mathbb{B}		& 	\mathbb{A}	\\
\end{array}
\right) 
\end{aligned}
\end{equation}
with
\begin{equation}
\begin{aligned}
\label{a}
\mathbb{A} =& 
\left(
\begin{array}{cc}
 \xi^H&	\xi_{12}	\\
 \xi_{12}& \xi^V	\\
\end{array}
\right)
\end{aligned}
\end{equation}
and
\begin{equation}
\begin{aligned}
\label{b}
\mathbb{B} =& 
\left(
\begin{array}{cc}
 \xi_c&	\xi_t	\\
 \xi_t& \xi_c	\\
\end{array}
\right).
\end{aligned}
\end{equation}
For the convenience of discussion, we have set the chemical potential $\mu$ to be zero.
Here we introduce a gauge transform 
$\eta({\bf k})=(d_{1\mathbf{k}}, d_{2\mathbf{k}}, d_{1\mathbf{k+Q}}, d_{2\mathbf{k+Q}})^T$ = $K\psi({\bf k})$
with
\begin{equation}
\begin{aligned}
\label{k}
K =&\frac{1}{\sqrt{2}} 
\left(
\begin{array}{cc}
 \mathds{1}		&	-\mathds{1}	\\
 -\mathds{1}	&	-\mathds{1}	\\
\end{array}
\right),
\end{aligned}
\end{equation}
where $\mathds{1}$ is a 2$\times$2 identity. The gauge transform $K$ satisfies $K^\dagger K=\mathds{1}_{4\times 4}$.
A little algebra leads to 
\begin{equation}
\begin{aligned}
	\mathcal{H}_0	&= \sum_{\bf k} \psi^\dagger({\bf k}) K^\dagger K \mathbb{W}_{\bf k}^\prime K^\dagger K \psi({\bf k})\\
	&= \sum_{\bf k} \eta^\dagger({\bf k}) K \mathbb{W}_{\bf k}^\prime K^{\dagger} \eta({\bf k})
\end{aligned}
\end{equation}
with
\begin{equation}
\begin{aligned}
\label{eq:kwk}
K^\dagger \mathbb{W}_{\bf k}^\prime K =&
\left(
\begin{array}{cc}
 \mathbb{A}+\mathbb{B}	&	0	\\
 0	&	\mathbb{A}-\mathbb{B}	\\
\end{array}
\right).
\end{aligned}
\end{equation}
Here $\mathbb{A}$ and $\mathbb{B}$ matrices follow $\mathbb{A}({\bf k})=\mathbb{A}({\bf k+Q})$ and $\mathbb{B}({\bf k})=-\mathbb{B}({\bf k+Q})$, respectively.
By carefully collecting terms in Eq.~(\ref{eq:kwk}), we confirm that
\begin{equation}
\begin{aligned}
\label{eq:wk2}
K^\dagger \mathbb{W}_{\bf k}^\prime K =&
\left(
\begin{array}{cc}
 \mathbb{M}_{\bf k}		&	0		\\
 0		&	\mathbb{M}_{\bf k+Q}	\\
\end{array}
\right)  = \mathbb{W}_{\bf k}^{\prime\prime}.
\end{aligned}
\end{equation}

By combining these results for the unitary transformations $U$ and $K$, we can map the 2-by-2 onto the 1-by-2 Hamiltonian:
$\mathbb{W}_{\bf k}^{\prime\prime} = K^\dagger \mathbb{W}_{\bf k}^\prime K =K^\dagger U\mathbb{W}_\mathbf{k} U^\dagger K$.


\begin{thebibliography}{99}

\bibitem{kamihara08} Y. Kamihara, T. Watanabe, M. Hirano, and H. Hosono, J. Am. Chem. Soc. \textbf{130}, 3296 (2008).

\bibitem{chenxh08} X. H. Chen, T. Wu, G. Wu, R. H. Liu, H. Chen, and D. F. Fang, Nature \textbf{453}, 761 (2008).

\bibitem{chengf08} G. F. Chen, Z. Li,D. Wu, G. Li, W. Z. Hu, J. Dong, P. Zheng, J. L. Luo, and N. L. Wang, Phys. Rev. Lett. \textbf{100}, 247002 (2008).

\bibitem{ren08} Z. A. Ren, W. Lu, J. Yang, W. Yi, X. L. Shen, C. Zheng, G. C. Che, L. X. Dong, L. L. Sun, Z. Fang, and Z. X. Zhao, Chin. Phys. Lett. \textbf{25}, 2215 (2008).

\bibitem{wang08} C. Wang, L. J. Li, S. Chi, Z. W. Zhu, Z. Ren, Y. Y. Li, Y. T. Wang, X. Lin, Y. K. Luo, S.Jiang,
X. F. Xu, G. H. Cao, and Z. A. Xu, Europhysics. Lett. \textbf{83}, 67006 (2008). 

\bibitem{pcdai08} Clarina de la Cruz, Q. Huang, J. W. Lynn, Jiying Li, W. Ratcliff II, J. L. Zarestky, H. A. Mook, G. F. Chen, J. L. Luo, N. L. Wang, Pengcheng Dai, Nature {\bf 453}, 899 (2008).

\bibitem{rotter08} M. Rotter, M. Tegel, and D. Johrendt, Phys. Rev. Lett. {\bf 101}, 107006 (2008).

\bibitem{ALeithe-Jasper:2008} A. Leither-Jasper, W. Schnelle, C. Geibel, and H. Rosner, Phys. Rev. Lett. {\bf 101}, 207004 (2008).

\bibitem{NNi:2008} N. Ni, S. L. Bud'ko, A. Kreyssig, S. Nandi, G. E. Rustan, A. I. Goldman, S. Gupta, J.D. Corbett, A. Kracher, and P. C. Canfield, Phys. Rev. B {\bf 78}, 014507 (2008).

\bibitem{HQLuo:2008} H. Q. Luo, Z. S. Wang, H. Yang, P. Cheng, X. Y. Zhu, and H. H. Wen, Supercond. Sci. Technol. {\bf 21}, 125014 (2008).

\bibitem{GFChen:2008} G. F. Chen, Z. Li, J. Dong, G. Li, W. Z. Hu, X. D. Zhang, X. H. Song, P. Zheng, N. L. Wang, and J. L. Luo, Phys. Rev. B {\bf 78}, 224512 (2008).

\bibitem{ASSefat:2008} A. S. Sefat, R. Y. Jin, M. A. McGuire, B. C. Sales, D. J. Singh, and D. Mandrus, Phys. Rev. Lett. {\bf 101}, 117004 (2008).

\bibitem{LJLi:2009} L. J. Li, Y. K. Luo, Q. B. Wang, H. Chen, Z. Ren, Q. Tao, Y. K. Li,
X. Lin, M. He, Z. W. Zhu, G. H. Cao, and Z. A. Xu, New J. Phys. {\bf 11}, 025008 (2009).

\bibitem{SRSaha:2009} S. R. Saha, N. P. Butch, K. Kirshenbaum, and J. Paglione, Phys. Rev. B {\bf 79}, 224519 92009).

\bibitem{gordon10} R. T. Gordon, H. Kim, N. Salovich, R. W. Giannetta, R. M. Fernandes, V. G. Kogan, T. Prozorov, S. L. Budko, P. C. Canfield, M. A. Tanatar, and R. Prozorov, Phys. Rev. B {\bf 82}, 054507 (2010).

\bibitem{MGVavilov:2009} M. G. Vavilov, A. V. Chubukov, and A. B. Vorontsov, Supercond. Sci. Technol. {\bf 23}, 054011 (2009).

\bibitem{TZhou:2011} T. Zhou, H. Huang, Y. Gao, J.-X. Zhu, and C. S. Ting, Phys. Rev. B {\bf 83}, 214502 (2011).

\bibitem{si09} Q. Si, Nat. Phys. {\bf 5}, 639 (2009).

\bibitem{hirschfeld11} P. J. Hirschfeld, M. M. Korshunov, and I. I. Mazin, Rep. Prog. Phys. {\bf 74}, 124508 (2011).

\bibitem{hu12} J. Hu and H. Ding, Sci. Rep. {\bf 2}, 381 (2012).

\bibitem{SBZhang:2011} S. B. Zhang, Y. F. Guo, J. J. Li, X. X. Wang, K. Yamaura, and E. Takayama-Muromachi, Physica C {\bf 471}, 600 (2011).

\bibitem{YQi:2011} Y. Qi, Z. Gao, L. Wang, X. Zhang, D. Wang, C. Yao, C. Wang, C. Wang, and Y. Ma, Europhys. Lett. {\bf 96}, 47005 (2011).

\bibitem{YNishikubo:2010} Y. Nishijubo, S. Kakiya, M. Danura, K. Kudo, and M. Nohara, J. Phys. Soc. Jpn. {\bf 79}, 095002 (2010).

\bibitem{FHan:2009} F. Han, X. Zhu, P. Ceng, G. Mu, Y. Jia, L. Fang, Y. Wang, H. Luo, B. Zeng, B. Shen, L. Shan, C. Ren, and H.-H. Wen, Phys. Rev. B {\bf 80}, 024506 (2009).

\bibitem{onari09} Seiichiro Onari and Hiroshi Kontani, Phys. Rev. Lett. {\bf 103}, 177001 (2009).

\bibitem{boyd10} G. R. Boyd, P. J. Hirschfeld, and T. P. Devereaux, Phys. Rev. B {\bf 82}, 134506 (2010).

\bibitem{efremov11} D. V. Efremov, M. M. Korshunov, O. V. Dolgov, A. A. Golubov, and P. J. Hirschfeld, Phys. Rev. B {\bf 84}, 180512(R) (2011).

\bibitem{wang13} Y. Wang, A. Kreisel, and P. J. Hirschfeld, Phys. Rev. B {\bf 87}, 094504 (2013).

\bibitem{fernandes12} R. M. Fernandes, M. G. Vavilov, and A. V. Chubukov, Phys. Rev. B {\bf 85}, 140512(R) (2012).

\bibitem{berlijn12} Tom Berlijn, Chia-Hui Lin, William Garber, and Wei Ku, Phys. Rev. Lett. {\bf 108}, 207003 (2012).

\bibitem{AVBalatsky:2006} A. V. Balatsky, I. Vekhter, and J.-X. Zhu, Rev. Mod. Phys. {\bf 78}, 373 (2006).

\bibitem{li09} Yuke Li, Xiao Lin, Qian Tao, Cao Wang, Tong Zhou, Linjun Li, Qingbo Wang, Mi He, Guanghan Cao and Zhu'an Xu, New J. Phys. {\bf 11}, 053008 (2009).

\bibitem{li10} Yuke Li, Jun Tong, Qian Tao, Chunmu Feng, Guanghan Cao, Weiqiang Chen, Fu-chun Zhang and Zhu-an Xu, New J. Phys. {\bf 12}, 083008 (2010).

\bibitem{yao12} Zi-Jian Yao, Wei-Qiang Chen, Yu-ke Li, Guang-han Cao, Hong-Min Jiang, Qian-En Wang, Zhu-an Xu, and Fu-Chun Zhang, Phys. Rev. B {\bf 86}, 184515
(2012).


\bibitem{cheng10} Peng Cheng, Bing Shen, Jiangping Hu, and Hai-Hu Wen, Phys. Rev. B {\bf 81} 174529 (2010).

\bibitem{li11} Jun Li, Yanfeng Guo, Shoubao Zhang, Shan Yu, Yoshihiro Tsujimoto, Hiroshi Kontani, Kazunari Yamaura, and Eiji Takayama-Muromachi, Phys. Rev. B {\bf 84}, 020513(R) (2011).

\bibitem{li12a} Jun Li, Yanfeng Guo, Shoubao Zhang, Yoshihiro Tsujimoto, Xia Wang, C.I. Sathish, Shan Yu, Kazunari Yamaura, Eiji Takayama-Muromachi, Solid State Commun. {\bf 152}, 671 (2012).

\bibitem{li12b} J. Li, Y. F. Guo, S. B. Zhang, J. Yuan, Y. Tsujimoto, X. Wang, C. I. Sathish, Y. Sun, S. Yu, W. Yi, K. Yamaura, E. Takayama-Muromachiu, Y. Shirako, M. Akaogi, and H. Kontani, Phys. Rev. B {\bf 85}, 214509 (2012).

\bibitem{christianson08} A. D. Christianson, E. A. Goremychkin, R. Osborn, S. Rosenkranz, M. D. Lumsden, C. D. Malliakas, I. S. Todorov, H. Claus, D. Y. Chung, M. G. Kanatzidis, R. I. Bewley, and T. Guidi, Nature {\bf456}, 930 (2008).

\bibitem{HDing:2008} H. Ding, P. Richard, K. Nakayama, K. Sugawara, T. Arakane, Y. Sekiba, A. Takayama, S. Souma, T. Sato, T. Takahashi,  Z. Wang, X. Dai, Z. Fang, G. F. Chen, J. L. Luo, and N. L. Wang, Europhys. Lett. {\bf 83}, 47001 (2008).

\bibitem{hanaguri10} T. Hanaguri, S. Niitaka, K. Kuroki, and H. Takagi, Science {\bf 328}, 474 (2010).

\bibitem{fernandes10} R. M. Fernandes and J. Schmalian, Phys. Rev. B {\bf 82}, 014521 (2010).

\bibitem{SRaghu:2008} S. Raghu, X.-L. Qi, C.-X. Liu, D. J. Scalapino, and S.-C. Zhang, Phys. Rev. B {\bf 77}, 220503 (2008).

\bibitem{YYTai:2013} Y.-Y. Tai, J.-X. Zhu, M. J. Graf, and C. S. Ting, arXiv:1303.1446 (To appear in Europhys. Lett.).

\bibitem{wadati10} H. Wadati, I. Elfimov, and G. A. Sawatzky, Phys. Rev. Lett. {\bf 105}, 157004 (2010).

\bibitem{Nakamura11} Kazuma Nakamura, Ryotaro Arita, and Hiroaki Ikeda, Phys. Rev. B {\bf 83}, 144512 (2011).

\bibitem{ideta13a} S. Ideta, T. Yoshida, M. Nakajima, W. Malaeb, T. Shimojima, K. Ishizaka, A. Fujimori, H. Kimigashira, K. Ono, K. Kihou, Y. Tomioka, C. H. Lee, A. Iyo, H. Eisaki, T. Ito, and S. Uchida, Phys. Rev. B {\bf  87}, 201110(R) (2013).


\bibitem{ideta13b} S. Ideta, T. Yoshida, M. Nakajima, W. Malaeb, T. Shimojima, K. Ishizaka, A. Fujimori, H. Kimigashira, K. Ono, K. Kihou, Y. Tomioka, C. H. Lee, A. Iyo, H. Eisaki, T. Ito, S. Uchida, arXiv:1304.5860v1.

\bibitem{MKano:2009} M. Kano, Y. Kohama, D. Graf, F. Balakirev, A. S. Sefat, M. A. Mcguire, B. C. Sales, D. Mandrus, and S. W. Tozer, J. Phys. Soc. Jpn. {\bf 78}, 084719 (2009).


\bibitem{abrikosov61} A. A. Abrikosov and L. P. Gor'kov, Sov. Phys. JETP {\bf 12}, 1243 (1961).

\bibitem{MFranz:1997} M. Franz, C. Kallin, A. J. Berlinksky, and M. I. Salkola, Phys. Rev. B {\bf 56}, 7882 (1997).

\bibitem{TDas:2011} T. Das, J.-X. Zhu, and M. J. Graf, Phys. Rev. B {\bf 84}, 134510 (2011).

\bibitem{KOhishi:2007} K. Ohishi, R.H . Heffner, G. D. Morris, E. D. Bauer, M. J. Graf, J.-X. Zhu, L. A. Morales, J. L. Sarrao, M. J. Fluss, D. E. MacLaughlin, L. Shu, W. Higemoto, and T. U. Ito, Phys. Rev. B {\bf 76}, 064504 (2007).


\bibitem{RBeaird:2012} R. Beaird, I. Vekhter, and J.-X. Zhu, Phys. Rev. B {\bf 86}, 140507 (2012).

\bibitem{PALee:2008} P. A. Lee and X.-G. Wang, Phys. Rev. B {\bf 78}, 144517 (2008).

\bibitem{CCao:2008} C. Cao, P. J. Hirschfeld, and H.-P. Cheng, Phys. Rev. B {\bf 77}, 220506(R) (2008).

\bibitem{KKuroki:2008} K. Kuroki, S. Onari, R. Arita, H. Usui, Y. Tanaka, H. Kontani, and H. Aoki, Phys. Rev. Lett. {\bf 101}, 087004 (2008).

\bibitem{YRan:2009} Y. Ran, F. Wang, H. Zhai, A. Vishwanath, and D.-H. Lee, Phys. Rev. B {\bf 79}, 014505 (2009).

\bibitem{DZhang:2009} D. Zhang, Phys. Rev. Lett. {\bf 103}, 186402 (2009). 

\bibitem{JHu:2012} J. Hu and N. Hao, Phys. Rev. X {\bf 2}, 021009 (2012).


\bibitem{mazin08} I. I. Mazin, D. J. Singh, M. D. Johannes, and M. H. Du, Phys. Rev. Lett. {\bf 101}, 057003 (2008).

\bibitem{wang09} Fa Wang, Hui Zhai, Ying Ran, Ashvin Vishwanath, and Dung-Hai Lee, Phys. Rev. Lett. {\bf 102}, 047005 (2009).

\bibitem{YBang:2009} Y. Bang,  Europhys. Lett. {\bf 86}, 47001 (2009).

\bibitem{IIMazin:2010} I. I. Mazin, Nature {\bf 464}, 183 (2010).

\bibitem{BNachumi:1996} B. Nachumi, A. Keren, K. Kojima, M. Larkin, G. M. Luke, J. Merrin, O. Tchernyshoev, Y. J. Uemura, N. Ichikawa, M. Goto, and S. Uchida, Phys. Rev. Lett. {\bf 77}, 5421 (1996).

\bibitem{Mackenzie1998} A. P. Mackenzie, 
R. K. W. Haselwimmer, A. W. Tyler, G. G. Lonzarich, Y. Mori, S. Nishizaki, and Y. Maeno,
Phys. Rev. Lett. {\bf  80}, 161 (1998).

\bibitem{Kycia1998} J. B. Kycia, J. I. Hong, 
M. J. Graf, J. A. Sauls,  D. N. Seidman, and W. P. Halperin, Phys. Rev. B {\bf 58}, R603 (1998).

\bibitem{Jutier2008} F. Jutier, 
 G. A. Ummarino, J.-C. Griveau, F. Wastin, E. Colineau, J. Rebizant, N. Magnani, and R. Caciuffo,
Phys. Rev. B {\bf 77}, 024521 (2008). 

\bibitem{KHashimoto:2009} K. Hashimoto, T. Shibauchi, S. Kasahara, K. Ikada, S. Tonegawa, T. Kato, R. Okazaki, C. J. van der Beek, M. Konczykowski, H. Takeya, K. Hirata, T. Terashima, and Y. Matsuda, Phys. Rev. Lett. {\bf 102}, 207001 (2009).

\bibitem{RTGordon:2009a} R. T. Gordon, N. Ni, C. Martin, M. A. Tanatar, M. D. Vannette, H. Kim, G. D. Samolyuk, J. Schmalian, S. Nandi, A. Kreyssig, A. I. Goldman, J. Q. Yan, S. L. Bud'ko, P. C. Canfield, and R. Prozorov, Phys. Rev. Lett. {\bf 102}, 127004 (2009).

\bibitem{RTGordon:2009b} R. T. Gordon, C. Martin, H. Kim, N. Ni, M. A. Tanatar, J. Schmalian, I. I. Mazin, S. L. BudÕko, P. C. Canfield, and R. Prozorov, Phys. Rev. B {\bf 79}, 100506(R) (2009).


\bibitem{HKim10} H. Kim, R. T. Gordon, M. A. Tanatar, J. Hua, U. Welp, W. K. Kwok, N. Ni, S. L. Bud'ko, P. C. Canfield, A. B. Vorontsov, and R. Prozorov, Phys. Rev. B {\bf 82}, 060518 (2010).

\bibitem{JYong:2011} J. Yong, S. Lee, J. Jiang, C. Bark, J. Weiss, E. Hellstrom, D. Larbalestier, C. Eom, and T. Lemberger, Phys. Rev. B {\bf 83}, 104510 (2011).

\bibitem{AABarannik:2011} A. A. Barannik, N. T. Cherpak, N. Ni, M. A. Tanatar, S.A. Vitusevich, V. N. Skresanov, P. C. Canfield, R. Prozorov, V.V. Glamazdin, and K. I. Torokhtii, Low Temp. Phys. {\bf 37}, 725 (2011).

\bibitem{TJWilliam:2009} T. J. Williams, A. A. Aczel, E. Baggio-Saitovitch, S. L. Bud'ko, P. C. Canfield, J. P. Carlo, T. Goko, J. Munevar, N. Ni, Y. J. Uemura, W. Yu, and G. M. Luke, Phys. Rev. B {\bf 80}, 094501 (2009).

\bibitem{HHuang:2012} Huaixiang Huang, Yi Gao, Jian-Xin Zhu, and C. S. Ting, Phys. Rev. Lett. {\bf 109}, 187007 (2012).

\bibitem{scalapino92} D. J. Scalapino, S. R. White, and S. C. Zhang, Phys. Rev. Lett. {\bf 68}, 2830 (1992).

\bibitem{scalapino93} D. J. Scalapino, S. R. White, and S. C. Zhang, Phys. Rev. B {\bf 47}, 7995 (1993).

\bibitem{bouadim11}  Karim Bouadim, Yen Lee Loh, Mohit Randeria and Nandini Trivedi, Nat. Phys. {\bf 7}, 884 (2011).


\bibitem{ghosal01} Amit Ghosal, Mohit Randeria, and Nandini Trivedi, Phys. Rev. B {\bf 65}, 014501 (2001).

\bibitem{benfatto04} L. Benfatto, A. Toschi, and S. Caprara, Phys. Rev. B {\bf 69}, 184510 (2004).

\bibitem{Ramakrishnan89} T. V. Ramakrishnan, Phys. Scr. {\bf T27}, 24 (1989).

\bibitem{ABVorontsov:2009}  A. B. Vorontsov, M. G. Vavilov, and A. V. Chubukov, Phys. Rev. B
{\bf 79}, 140507 (2009).

\bibitem{Hashimoto09} K. Hashimoto, T. Shibauchi, S. Kasahara, K. Ikada, S. Tonegawa, T. Kato, R. Okazaki, C. J. van der Beek, M. Konczykowski, H. Takeya, K. Hirata, T. Terashima, and Y. Matsuda, Phys. Rev. Lett. {\bf 102}, 207001 (2009).

\bibitem{CMartin:2009} C. Martin, R. T. Gordon, M. A. Tanatar, H. Kim, N. Ni, S. L. Bud'ko, P. C. Canfield, H. Luo, H. H. Wen, Z. Wang, A. B. Vorontsov, V. G. Kogan, and R. Prozorov, Phys. Rev. B {\bf 80}, 020501(R) (2009).

\bibitem{JKim:2012}  Jeehoon Kim, N. Haberkorn, M. J. Graf, I. Usov, F. Ronning, L. Civale, E. Nazaretski, G. F. Chen, W. Yu, J. D. Thompson, and R. Movshovich, Phys. Rev. B {\bf 86}, 144509 (2012).

\bibitem{JHXu:1995} J. H. Xu, J. H. Miller, Jr., and C. S. Ting, Phys. Rev. B {\bf 51}, 424
(1995).

\end{thebibliography}
\end{document}